\documentclass{article}
\usepackage[utf8]{inputenc}
\usepackage{biblatex}
\usepackage{todonotes}
\usepackage{graphics}
\usepackage{subcaption}
\PassOptionsToPackage{hyphens}{url}
\usepackage{hyperref}

\usepackage{draftwatermark}

\title{Epidemic? The Attack Surface of German Hospitals during the COVID-19 Pandemic}

\author{
  Johannes Klick\\
  Alpha Strike Labs\\
  \texttt{j.klick@alphastrike.io}
  \and
  Robert Koch\\
  Universität der Bundeswehr\\
  \texttt{robert.koch@unibw.de}
  \and
  Thomas Brandstetter\\
  Limes Security/FHSTP\\
  \texttt{tbr@limessecurity.com}
}
\date{January 2021}
\nocite{*}

\SetWatermarkText{\textbf{Preprint Version}\\ Paper submitted to Peer Review Process}
\SetWatermarkLightness{0.85}
\SetWatermarkScale{1.8}

\begin{document}

\maketitle

\begin{abstract}
    In our paper we analyze the attack surface of German hospitals and healthcare providers in 2020 during the COVID-19 Pandemic. The analysis looked at the publicly visible attack surface utilizing a Distributed Cyber Recon System, utilizing distributed Internet scanning, Big Data methods and scan data of 1,483 GB from more than 89 different global Internet scans. From the 1,555 identified German clinical entities, security posture analysis was conducted by looking at more than 13,000 service banners for version identification and subsequent CVE-based vulnerability identification. Primary analysis shows that 32 percent of the analyzed services were determined as vulnerable to various degrees and 36 percent of all hospitals showed numerous  vulnerabilities. Further resulting vulnerability statistics were mapped against size of organization and hospital bed count.

\end{abstract}

\section{Introduction}
\label{sec:intro}

In October 2020, US-CERT had issued a warning regarding the increasing ransomware activity in the healthcare sector~\cite{triagencyalert}.
It was common knowledge~\cite{wiredhospitalransomware} that healthcare organizations are promising targets for ransomware gangs. Surprisingly, in the very beginning of the COVID-19 pandemic, several ransomware gangs actually pledged not to hit hospitals because of the ongoing scourge.
The Maze and DoppelPaymer groups, for instance, said they would not target healthcare facilities and, if accidentally hit, would provide the decryption keys at no charge.
As another example, the Netwalker operators, stated they would not intentionally target hospitals, however if accidentally hit, the hospital would still have to pay the ransom.
Other attacker groups had much less scruples.
In the end unfortunately, ransomware incidents against hospitals skyrocketed in October 2020, with the strongest  surge against those victims was shown by weaponizing the Ryuk ransomware against 250 U.S.-based hospitals and clinics~\cite{wiredransomwave}.
The criticalness of the ransomware attack wave against the U.S. was demonstrated by the mentioned, very rare tri-agency ransomware alert issued by the Federal Bureau of Investigation (FBI), U.S. Department of Health and Human Services (HHS), and Cybersecurity and Infrastructure Security Agency (CISA).
\newline

In an increasingly digitized and interconnected world however, those issues are of course not limited to the United States: In Germany in 2020, there was an intense discussion about an incident involving the death of a patient who had to be taken to a distant hospital, because the closest hospital was logged out of emergency treatment due to an ongoing ransomware attack (e.g. see \cite{wiredduesseldorf}).
Even though it was the patient's critical health condition that was ultimately determined as the actual cause of death, and not the longer time it took to get to the more distant clinic, this specific example underscores the increasing threats posed by cyber-attacks, particularly in the healthcare sector.
\newline

It must be noted however, that cybersecurity threats in the healthcare and medical sector are anything but new.
On the one hand, healthcare and medical production has always been an innovative field, in which new procedures and technologies are used. On the other hand, long life cycles or rather the long service life of products in this area as well as the need for time-consuming re-certifications when e.g. changing or patching the software, are known challenges. 
The need for comprehensive quality control and certification, especially in the medical field, is illustrated by the example of Therac-25 and the fatal incidents involving the faulty irradiation of patients already in the 1980s~\cite{lim1998engineering}.
Although the healthcare equipment of several vendors have a high security level nowadays, 
quite a large number of healthcare components and systems still have numerous security issues, some of them even being critical according to the Common Vulnerability Scoring System (CVSS)~\cite{CVSS}.
To add to that even worse, the attack surface stemming from complex healthcare networks and equipment is increasingly challenging~\cite{Forescout2020}.
\newline

With the aforementioned increase in cybercrime, this raises the question of what the posture of cybersecurity in the healthcare sector is, which weaknesses and vulnerabilities can be identified in the healthcare system, and what recommendations for action need to be derived.
In the very current context of the COVID-19 pandemic, we therefore chose to examine the cyber attack surface and vulnerability posture of hospitals and clinical providers in our home country Germany.

\section{Related Work}
\label{sec:related}

On the basis of innovation and (at that time generally) low security standards, the original birth of the very first piece of ransomware surprisingly took place also in the medical sector: 
In 1989 the malware "PC Cyborg", commonly also known as "AIDS Trojan"~\cite{Virus90} was distributed to probably approximately 20,000 people, among others the participants of a WHO conference on AIDS.
Under the guise of evaluation software, the first encryption Trojan was hidden, which was attributed to the American biologist Dr.~Joseph Popp.
Interestingly enough, the effects of the Trojan in the event of non-payment of the ransom were stored in the user agreements to be accepted by the user.\\

Despite this early appearance of this type of malware in computer history, "success" of ransomware failed to materialize for a long time.
Different ransomware variants such as "Fake Antivirus" (2001), GPCoder (2005), CRYZIP (2006) or QiaoZhaz (2007) appeared from 2001 onwards, but the attacks were still limited, mainly because of various technical or money-logistic reasons.
Some creative approaches like WinLock used SMS and phone calls to premium numbers, for example, to monetize attacks but a noteworthy crime-breakthrough came with CryptoLocker in 2013, introducing payments via Bitcoin.
While CryptoLocker was taken down in June 2014, it was the blueprint for numerous copycats as it showed that it was possible to earn millions within a few weeks.
Therefore, the right combination of public-key cryptography, the digital currency Bitcoin, anonymization possibilities by using the Tor network and providing a reliable decryption opened up a new business model which nowadays costs billions every year.
For example, \cite{RansomHist} gives an overview of the history and other details.\\




On the basis of various technical developments and improvements, it was thus possible for criminals to implement an effective digital blackmail model by means of simple cryptography, anonymous communication and simple, quasi-anonymous payment options.
Even if there are cases in which the data of the attacked system has been destroyed and actual recovery was never intended (for example, because no required key material was kept), these are in relation exceptions and stem either from errors in the technical implementation, or just from the attacker having other intentions than demanding ransom.
The success was based on the fact that victims who choose to pay mostly have a good chance of recovering their data; the attackers are thus motivated to enable correct decryption in order to keep their business model alive and thriving.\\

While in the earlier days the attackers chose victims randomly, which often were individuals  allowing only small money claims, over time and with increasing professionalization, much larger organizations were targeted and the attacks and claims became bolder.
In addition to the improvement of attack methods, the development of attackers is recently also characterized by a much more targeted approach~\cite{ransomoperations}.
The current focus is on companies and larger institutions where higher und thus more lucrative ransoms can be obtained.
Companies active in the grey area, which sell vulnerabilities and even 0days, extend this threat.
An example for this is the "MedPack" of the company GLEG Ltd. which contains 0days especially for the area of medical software~\cite{GLEG}.\\

The amount of the ransom is for obvious reasons based on a corresponding analysis of the target.
The blackmailers are also increasing the pressure on the victims by threatening to publish stolen data of the company, which has already happened several times~\cite{healthitsecurity2020}.\\

In theory, this trend can only be broken if no more payments are made over a longer period of time.
The technical prerequisite and basis for this are regular offline backups, as well as regular tests of the disaster recovery procedures, with dedicated checks on ransomware recovery.\\

Unfortunately, there are often worlds apart between theory and practice - backups are either not current and up-to-date or just not available due to misconfiguration, maybe because they are not kept offline and also encrypted, or critical aspects of the recovery process fails, because they have never been validated in the current environment.\\

Driven by the increase in ransomware attacks, companies are either considering investing in cyber-insurances in order to cushion possible financial damages and, in case of doubt, simply paying the ransom.
The statistics are telling: Already over 40 percent of cyber-insurance claims involve ransomware~\cite{zdnetransomclaims}.
Accordingly, some countries are considering banning the payment of ransoms already, in order to remove the basis for the business model.
The U.S. Department of the Treasury is already pointing out that ransom payments to groups or organizations on the sanctions list are punishable if they are not approved~\cite{treasuryransomapproval} by the Office of Foreign Assets Control (OFAC). Cyber-related Sanctions is a special section on the U.S. Department of the Treasury's website.\\

The difficulty of implementing this requirement in practice is, however, already evident in existing examples such as two police departments in Swansea, Massachusetts~\cite{swanseapoliceransom} and in Dickson, Tennessee: These departments, infected by the ransomware CryptoWall 2.0, have paid a ransom to recover their data.
With this background, it is a worthwhile question to explore the attack surface and security posture in the healthcare sector.\\






Against the background of the increasing ransomware campaigns, the outstanding importance of a functioning healthcare system, especially in the prevailing COVID-19 pandemic, and possible influencing factors through the short-term provision and integration of remote access and teleworking possibilities, we have conducted an in-depth investigation of the cyber attack surface of German hospitals based on the DIVI intensive register~\cite{diviregister}.\\

The rest of the paper is structured as follows: The current chapter gives context - it describes related work and background information on ransomware attacks in the healthcare sector as well as the overall development of this problem.
Chapter \ref{chap:DCS} describes the technical infrastructure that made our analysis possible: We describe our Distributed Cyber Recon System and how we used and extended it through our analysis.
In Chapter \ref{chap:Methodology} our methodology for attack surface detection of hospitals and clinical providers is presented, how we approached this from a healthcare entity identification point as well as attack surface correlation point of view.
Chapter \ref{chap:data} contains the data section where we describe the results of our finding in detail both verbally as through visualizations.
Chapter \ref{chap:discussion} describes a summary of the results of our analysis.






\section{Introduction to the Distributed Cyber Recon System (DCS)}
\label{chap:DCS}

The previous chapters have illustrated that it is both possible and feasible to attack hospitals and medical devices. However, the question arises how large the potential cyber attack surface of critical infrastructures like hospitals actually is?\\

Reconnaissance and, in particular, aggregation and representation of an organization in  
cyberspace is a major challenge. For this reason, there was a need for a novel search engine that can search the entire Internet (2.8 billion routed IP addresses) in a few hours for a network service or services/servers with a specific vulnerability in a matter of hours, allowing also mapping to a specific target organization.\\

In our Distributed Cyber Recon System (DCS) that was developed specifically for various recon and analysis tasks, we can answer questions like: What is the security posture of a specific organization? What is the attack surface of an entire group of organizations? Which systems belong to which organization in the first place?
Since plain Internet scan data is not sufficient, the scan data is augmented with additional information  such as Whois data, IP prefix, Autonomous System (AS) information, certificate information and geo-information about the IP of the system. The combination of this information in a Big Data approach enables a quite accurate representation of cyberspace.\\

For example, not only can the  distribution of selected system versions of a particular network service in an organization, or all detected Industrial Control Systems (ICS) be displayed on a map, but also systems organized by specific country.
IP prefix and IP ownership information can also be selected and aggregated using dynamic charts. This allows a recon analyst to get a quick overview of the cyber infrastructures of their own, as well as those of foreign  states, organizations and companies \cite{CCCamp2019}.\\

In our case, this DCS was used to analyze the security posture and publicly visible system attack surface of hospitals located in Germany. In the following passages, the methodology of our data collection or that of the DCS is explained in more detail.\\

\begin{figure}
    \centering
\includegraphics[width=1\textwidth]{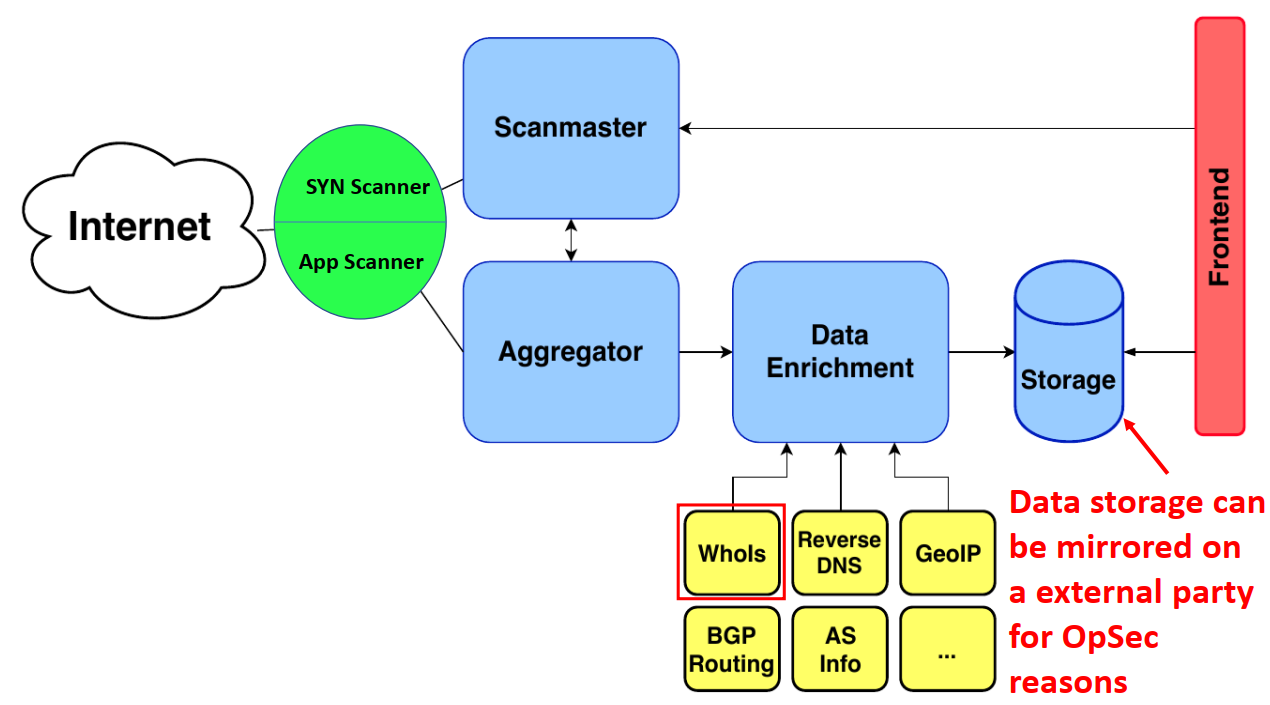}
\caption{Distributed Cyber Recon System Architecture.}
\label{pic:DCS_Structure}
\end{figure}

The DCS primarily consists of the search nodes, a backend and a frontend. The relationships between the individual components are shown in Figure \ref{pic:DCS_Structure}. The frontend is used by an analyst for operation setup and data analysis. The IPv4 network ranges, protocols, ports and scan algorithms to be scanned are defined in the frontend. The IPv4 range to be scanned is then pseudorandomized by the scan master in accordance with the selected scan algorithm, divided into several work units and distributed to the various scan nodes. The scan nodes are distributed worldwide for OPSEC as well as quality and correlation reasons and have different scan bandwidths.\\

With our DCS, it is therefore also possible to scan the same target areas simultaneously from different strategically interesting locations (e.g. different countries) as site groups and to compare the results. Also, if a scan node fails, the scan master will automatically detect this. Subsequently, the scan master will assign the work unit of the failed node to a new search node.
This ensures that all required IP addresses are always scanned, guaranteeing that the system produces consistent data. 
Experience has shown that result quality can significantly improve with an globally distributed group of scan nodes, as not all destinations are visible from all parts of the Internet due to various national or regional filtering approaches.\\

The search nodes consist of two primary components. First of all, the SYN scanner is active, which only sends TCP SYN or UDP packets. During the sending process, the last used destination addresses are stored in a ring buffer. At the same time, it is waiting for incoming TCP SYN ACK packets or UDP responses whose senders correspond to the destination addresses of the ring buffer. This prevents the search engine from being used as a DoS amplifier. Furthermore, the \emph{search nodes use more than 1,024 different IPv4 originator addresses} and can thus distribute the scan traffic. This allows it to stay below the radar of many intrusion detection systems and thus increases the scan data quality significantly.\\ 

As soon as a valid packet arrives from a destination address, the application scanner is started. The application scanner supports more than 60 different protocols and establishes full application connections with the goal of reading as much identification information as possible from the system.
Most of the protocols were implemented by ourselves, for several standard protocols the Zgrab implementation was used \cite{zgrab}.\\
After processing the IPv4 addresses of a work unit, the scan results are sent to the aggregator. The aggregator collects all results from search nodes and checks them for consistency. Then the data is enriched with other open sources of information in JSON format.\\

For example, the IPv4 scan data is enriched with the INETNUM and WHOIS information from the RIRS (RIPE, ARIN, AfriNic etc). Possible inconsistencies within the databases, e.g. overlapping prefixes, are resolved according to a self-developed method defined in \cite{klick2016}. In addition, the BGP data valid at the respective time is stored for each IPv4 address. This includes the BGP prefix annotated at the time, including all available autonomous system data. As a data source for the BGP information, the data of the Cooperative Association for Internet Data Analysis (CAIDA)~\cite{caida2020} is merged and processed.
In addition, reverse DNS records and Geo-IP information are added to each discovered active IPv4 address of the respective scan.
All data is stored in a NOSQL-based ElasticSearch database, which can be duplicated as an on-prem solution for discretionary analysis at any time.\\  

For the analysis of the hospital data, a separate subfrontend called Inspector was developed, to make this complex task for our human analysts more convenient and doable. The Inspector  only receives the names of the hospitals and the respective domain as input. Subsequently, all relevant entries in the database, such as the Whois Description field, or the common names of the collected TLS certificate information or the atomic system data, are analyzed for membership of the respective target set using advanced Big Data algorithms. 
In parallel, all subdomains of the added domains are searched. This is done by special best guess algorithms or by searching known certificate databases such as crt.sh. The Inspector had to be created as our analysts had to take a huge list of potential healthcare target organizations into account.\\ 

The final step is about vulnerability detection: After all network services of the defined reconnaissance targets, in our case hospitals and other healthcare providers, had been identified, the system descriptions or version strings read out were compared with the National Vulnerability Database (NVD) of NIST~\cite{NVD}. Through this step, all potential known vulnerabilities in detected software systems are identified.







\section{Methodology Attack Surface Detection of\\ Hospitals and Clinical Providers}
\label{chap:Methodology}

For the identification of the attack surface of the German hospitals, the German hospitals themselves had to be identified first. Therefore, we chose as a starting point the German DIVI registry\footnote{\url{https://www.intensivregister.de/\#/index}}, which was first established by the COVID-19 pandemic.\\

The DIVI Intensive Care Register records the free and occupied treatment capacities in intensive care medicine of about 1,300 acute hospitals in Germany on a daily basis.
During the pandemic and beyond, the registry makes it possible to identify bottlenecks in intensive medical care in a regional and temporal comparison. Thus, the DIVI Intensive Care Registry creates a valuable basis for response and data-driven action control in real time since April 2020.\\

From an approach perspective, we did the following: We extracted over 1,300 names of German hospitals with COVID-19 intensive care units from the DIVI Register. We then manually searched for the main website or domain of the corresponding hospital names and added them to the DIVI Registry data.\\

In the next processing step, the names and domain information were entered into the Inspector. The Inspector then analyzed a total of 89 different port/protocol scans.
A whopping amount of 1,483 GB  of data was analyzed on a system with 1 TB Ram, 64 CPU cores and 40 TB SSD storage and 72 TB HDD storage. The total computing time of the whole system was about 16 hours.\\  

Table \ref{tab:services} on page \pageref{tab:services} is a listing of the port/protocol combinations for which global scans for about 2.8 billion routed IPv4 addresses have been conducted.\\

\begin{table}[h!]
\begin{tabular}{ l c c r }
http-1000 & bacnet-47808 & postgres-5432 & openport-1025 \\
http-5985 &	bigip-443 &	qnapvuln-8080 &	openport-111\\
http-7547 &	cve20205902-443 &	redis-6379 &	openport-11211\\
http-80 &	dnp3-20000 &	s7-102 &	openport-11711\\
http-8008 &	imap-143 &	samba-445 &	openport-1201\\
http-8080 &	ipmi-623 &	snmpv1-161 &	openport-135\\
http-8088 &	ipp-631 &	snmpv2-161 &	openport-139\\
http-8888 &	kibana-5601 &	ssh-22 &	openport-1433\\
https-1443 & knx-3671 &	ssh-22022 &	openport-1521\\
https-443 &	ldap-389 &	ssh-2222 &	openport-1720\\
https-4433 & ldapudp-389 &	sworionrest-17778 &	openport-1723\\
https-4434 & modbus-502 &	telnet-23 &	openport-199\\
https-4444 & mongodb-27017 &	telnet-2323 &	openport-2012\\
https-5986 & mssqludp-1433 &	telnet-4786 &	openport-27017\\
https-8443 & mssqludp-1434 &	telnet-5938 &	openport-3306\\
dnstcp-53 &	mysql-3306 &	telnet-7070 &	openport-3389\\
elastic-9200 &	netbios-137 &	upnp-1900 &	openport-445\\
eniptcp-44818 &	ntp-123 &	winrm-5984 &	openport-469\\
fox-1911 & oracledb-1521 &	openport-993 &	openport-5037\\
ftp-21 & pop3-110 &	openport-995 &	openport-5432\\
openport-873 &	openport-6379 &	openport-5900 &	openport-5555\\
openport-9200 &	openport-8009 &	openport-5984 & openport-5601\\
openport-587 &	 		 & &\\
\end{tabular}
\caption{Scanned TCP and UDP ports during attack surface mapping.}
\label{tab:services}
\end{table}

After identification of the associated network services based on the certificate information, Whois and BGP/AS data, as well as the extended detection of subdomains, additional information about other hospitals could be collected.
For example, the cryptographic TLS certificate of hospital A might also include the domain of another hospital B of the same provider. Furthermore, generic search terms such as hospital, clinic, etc. were also added. In addition, the results were manually searched and any false positives were eliminated. Through this approach, the analysis of 1,300 hospitals of the DIVI registry could be extended to 1,555 hospitals.  

\newpage
\section{Data Section}
\label{chap:data}

\begin{figure}[th!]
\begin{center}
\includegraphics[width=0.80\linewidth]{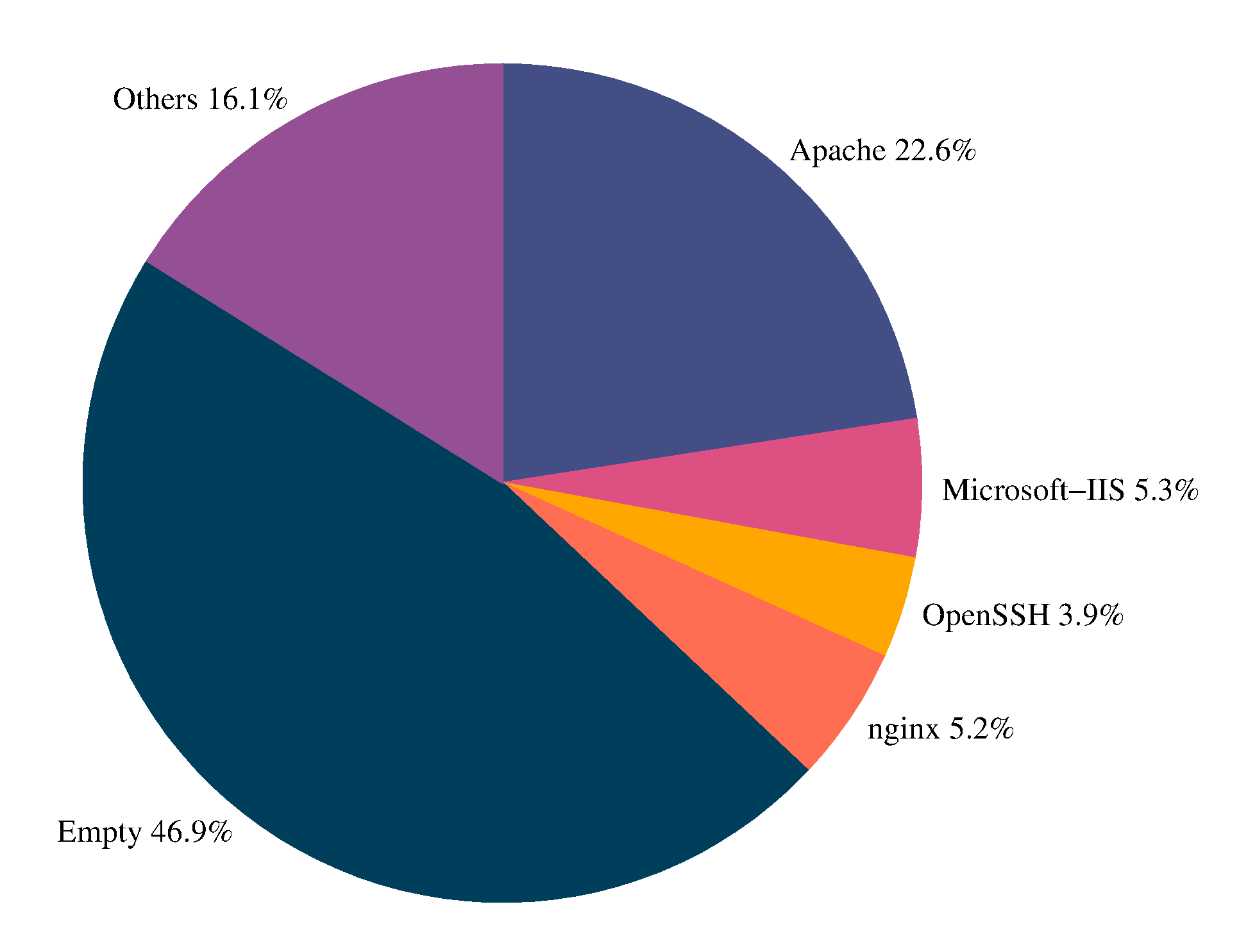}
\caption{This pie graph shows the distribution of the most common detected service banner grouped by major service application.}
\label{fig:krankenhaus-service}
\end{center}
\end{figure}

Our analysis of the 1,555 German hospitals revealed a digital attack surface of 13,497 network services, or 8.7 network services per hospital on average.
Figure \ref{fig:krankenhaus-service} shows the distribution of the main service banner groups of all identified hospital network services. Approximately 47 percent of all collected service banners are empty and thus comply with the common best-practice approach of not disclosing any software version information via service banner. This approach is very important because it makes it more difficult for attackers to identify the software used. This makes it subsequently harder for a potential attacker to determine the proper exploit/malware to use in an attack attempt. This is especially true for the use of automated attack scripts, often used by automated botnets.\\ 

We identified 1,228 hospitals and hospital operating companies that had network services that could be directly located. Approximately 300 other hospitals had no network services of their own, but only those that could be assigned to joint operating companies. However, since we do not know how the networks of the  joint hospital operating companies are related to the hospitals, we consider the whole operating company as a single hospital. Thus, we technically analyze 1,228 hospital entities and operating companies representing up to 1,555 different hospitals.
Of the 1,228 hospitals, 447 had vulnerable network services. This means that 36.4\% of all identified hospitals have vulnerabilities.

\begin{figure}[ht!]
  \centering
  \includegraphics[width=0.95\linewidth]{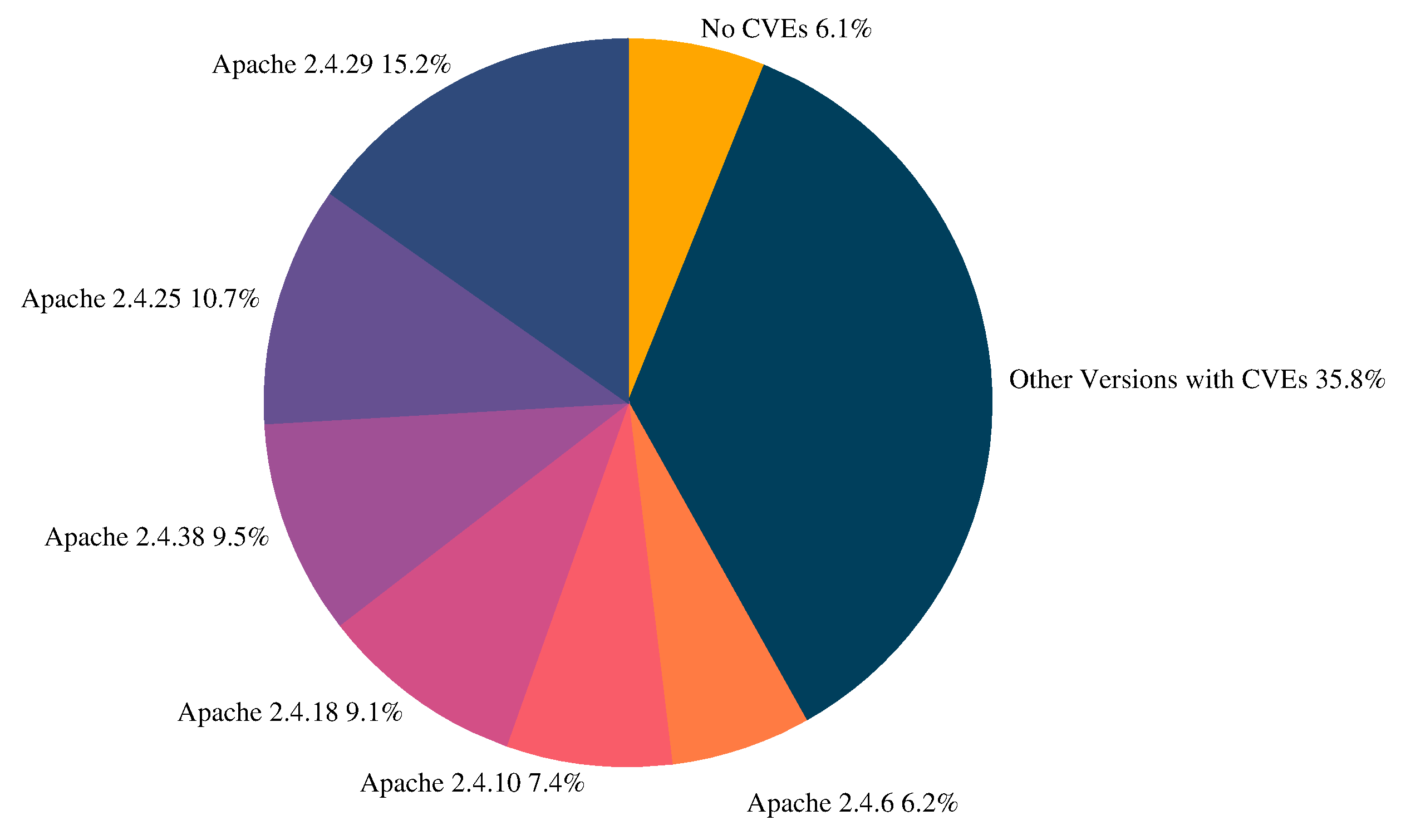}
  \caption{Version distribution of detected Apache web servers, roughly one third having known vulnerabilities. Note, that 2092 (68.43\%) Apache servers resulted in an undefined version and are not included.}
  \label{apache}
  \vspace*{\floatsep}
  \includegraphics[width=0.95\linewidth]{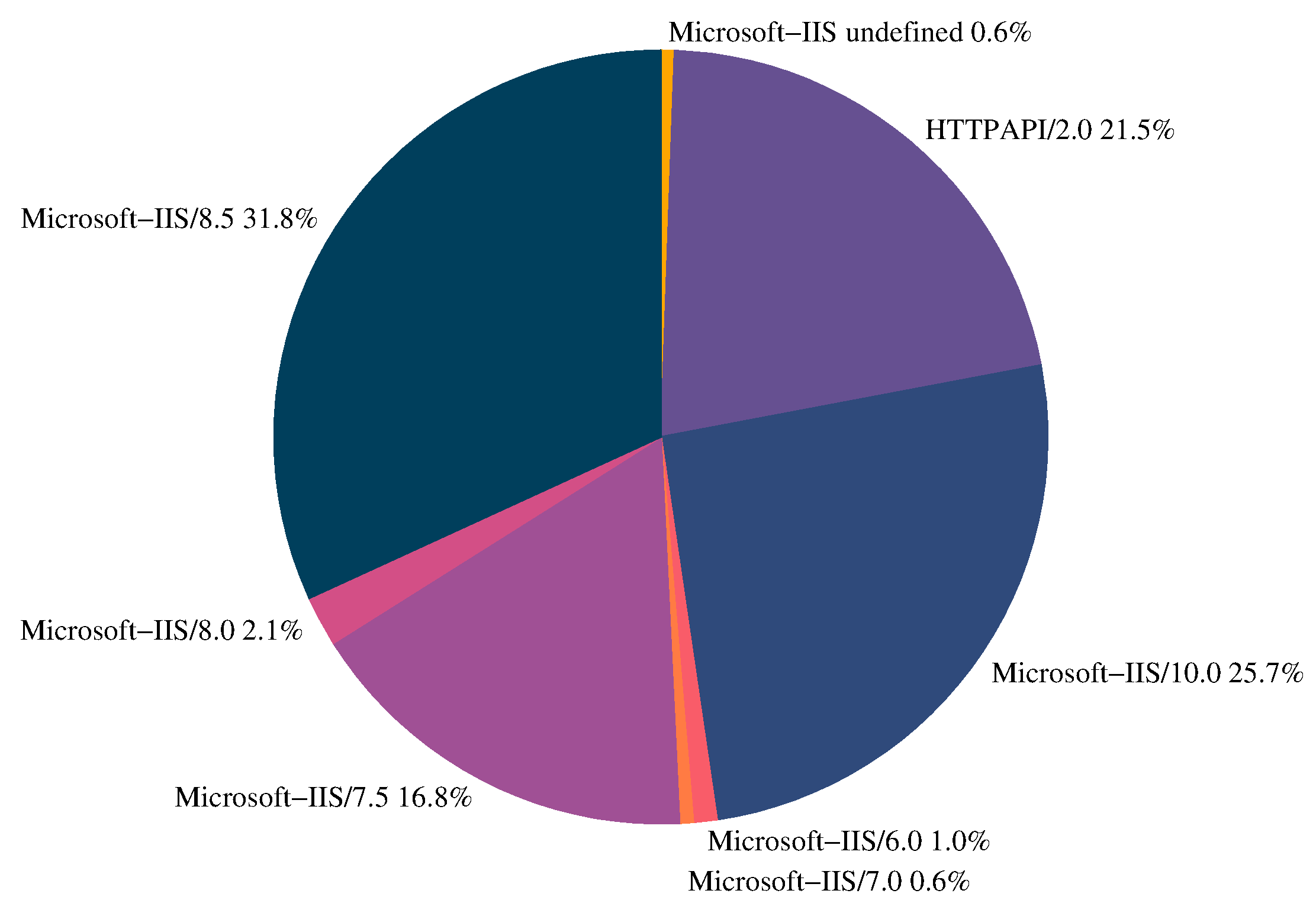}
  \caption{Distribution of detected versions of Microsoft Internet Information Services (IIS) webserver, indicating current as well as end-of-support versions in operation.}
  \label{iis}
\end{figure}
\newpage

\begin{figure}[htb!]
\begin{center}
\includegraphics[width=0.95\linewidth]{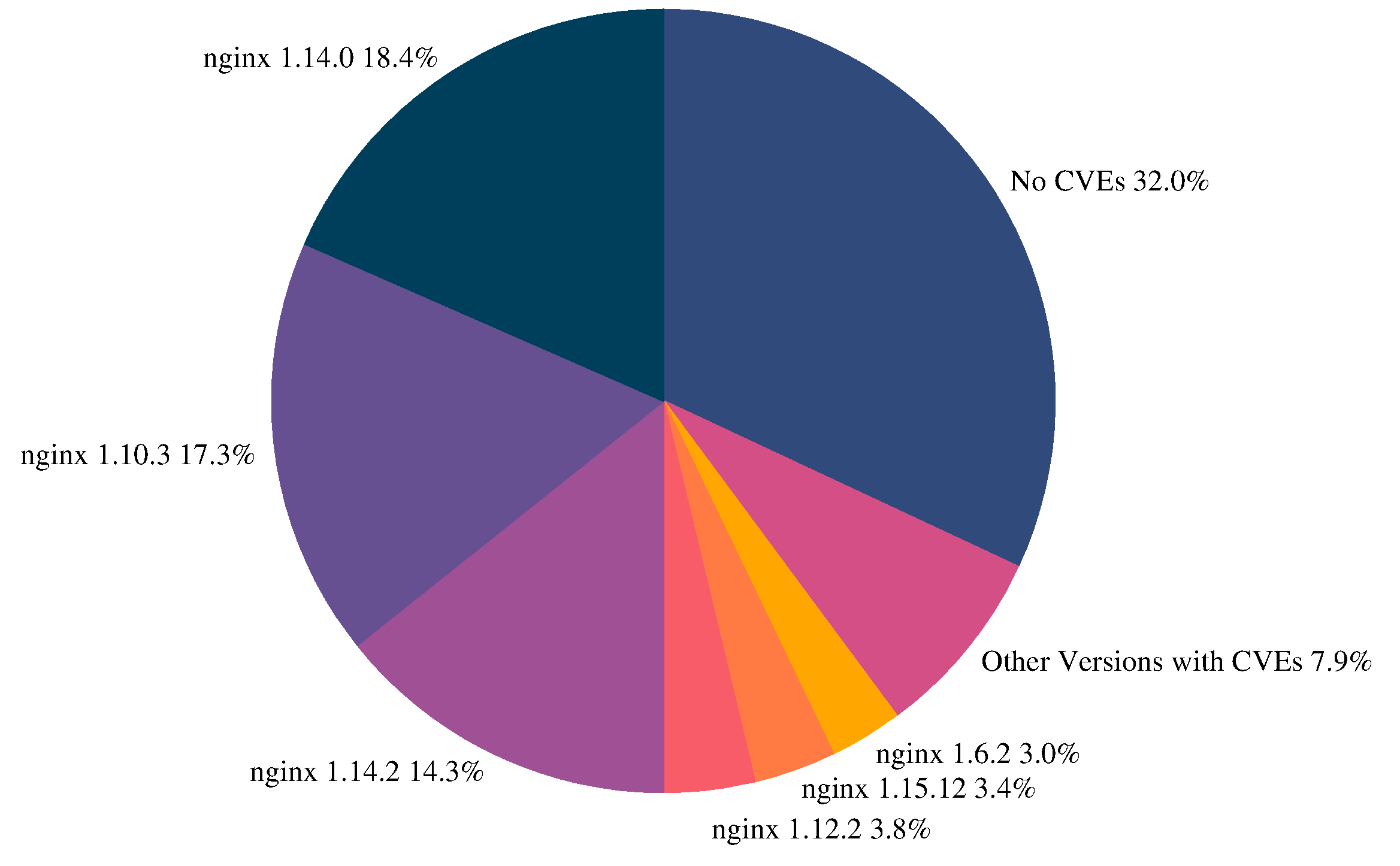}
\caption{Distribution of detected versions of nginx webserver, indicating current as well as end-of-support versions in operation. Note, that 444 (62.62\%) nginx servers resulted in an undefined version and are not included.}
\label{nginx}
\end{center}
\end{figure}

Figures \ref{apache}, \ref{iis} and \ref{nginx} show the version distribution of the three most common web servers Apache httpd, Microsoft IIS and nginx.
A well-known problem that exists both in the industrial as well as to a certain degree also in the healthcare sector became visible quite early in our analysis: Outdated services for which end-of-support had been announced already. The most noteworthy candidates we identified were e.g. \emph{Apache httpd version 2.2.x}, which became end-of-support in 12/2017 or \emph{Microsoft Internet Information Services 6.0} which became end-of-support in June 2015. It is unclear however why we found those legacy services on Internet-facing systems, as the issue of patch and update difficulty usually mainly affects internal medical components, not Internet infrastructure.\\



\begin{figure}[htb!]
  \begin{subfigure}[b]{0.5\linewidth}
    \centering
    \includegraphics[width=0.95\linewidth]{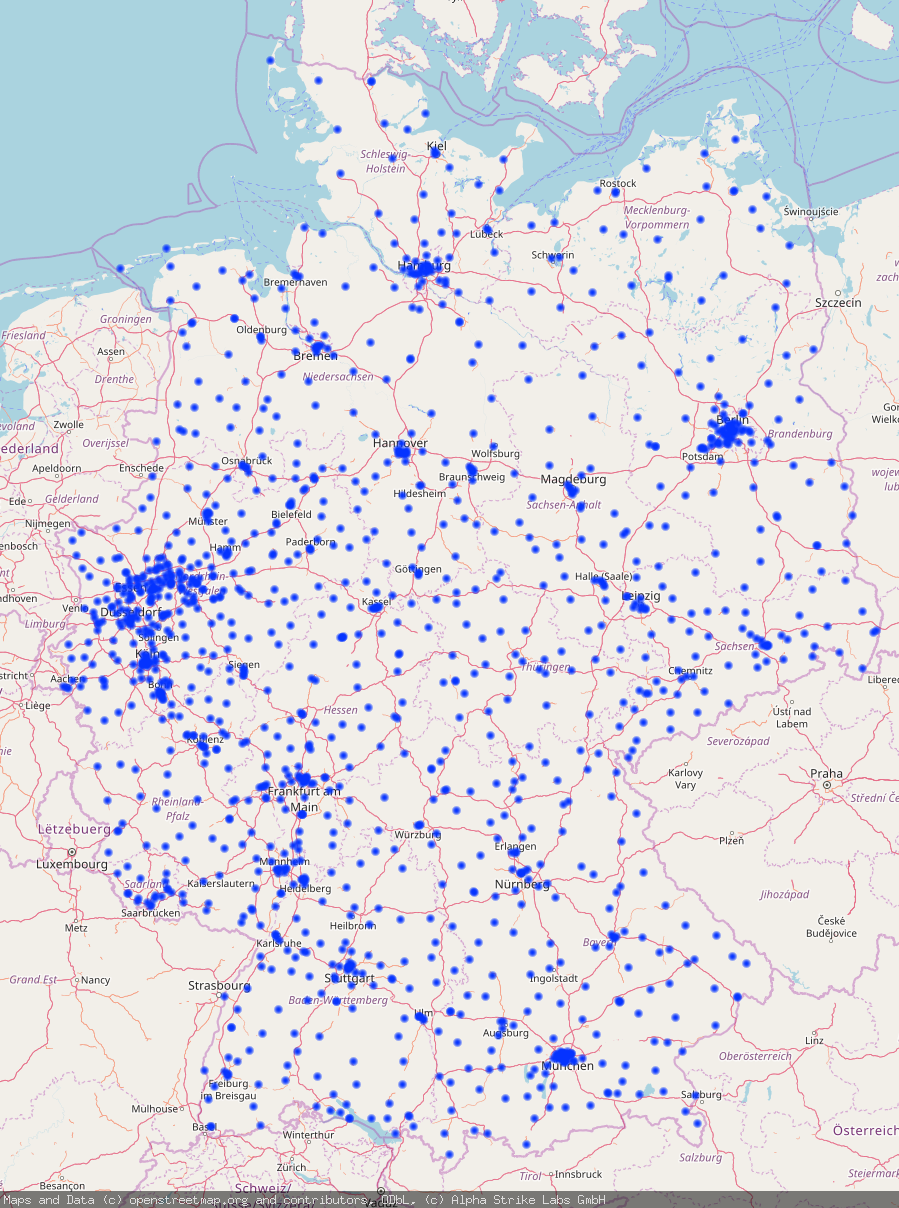} 
    \caption{Identified hospitals and geolocation according to the DIVI registry.} 
    \label{fig:divi}
    \vspace{4ex}
  \end{subfigure}
  \begin{subfigure}[b]{0.5\linewidth}
    \centering
    \includegraphics[width=0.95\linewidth]{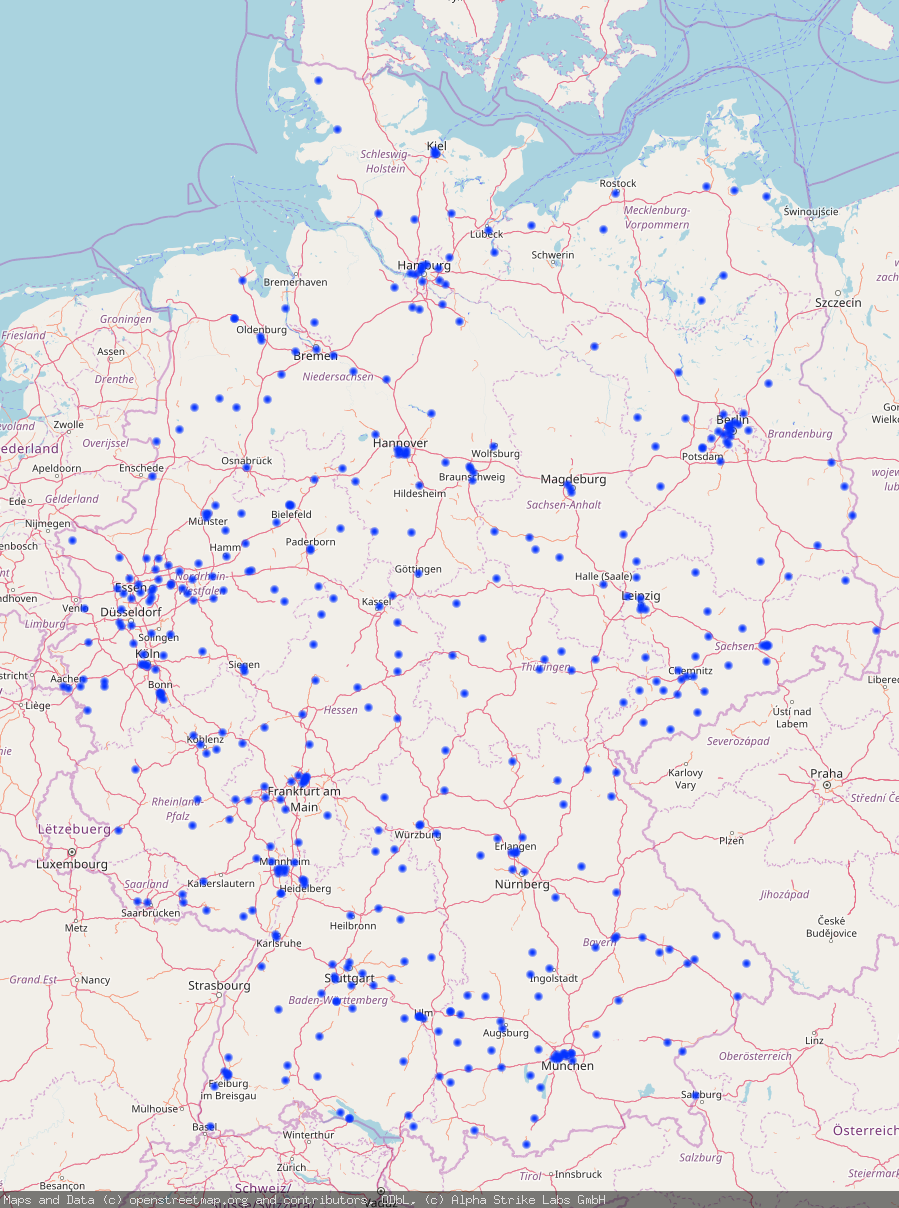}
    \caption{Identified DIVI hospitals having vulnerabilities.}
    \label{fig:divi-vuln} 
    \vspace{4ex}
  \end{subfigure} 
  \begin{subfigure}[b]{0.5\linewidth}
    \centering
    \includegraphics[width=0.95\linewidth]{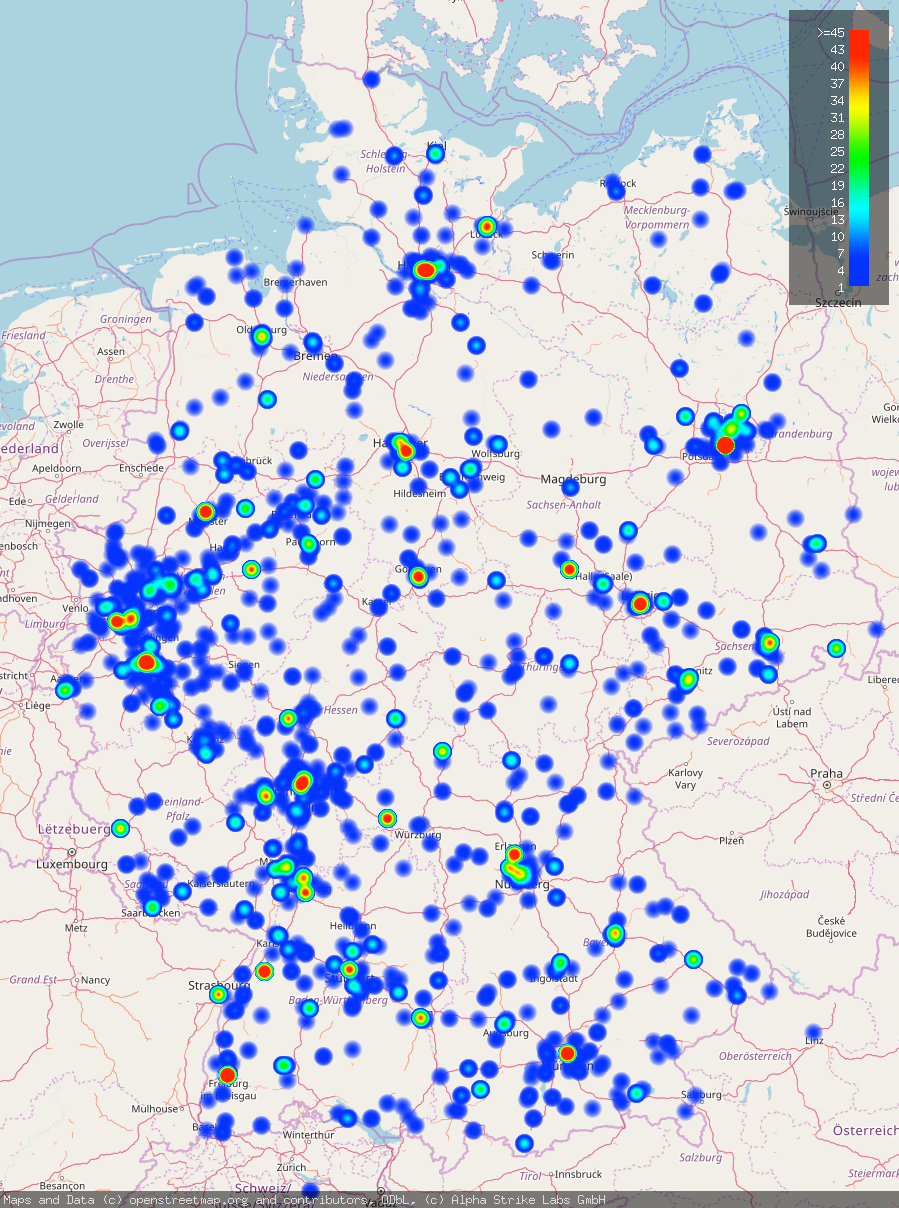} 
    \caption{All network services identified and the approximate Geo-IP location as heatmap.} 
    \label{fig:geoIP} 
  \end{subfigure}
  \begin{subfigure}[b]{0.5\linewidth}
    \centering
    \includegraphics[width=0.95\linewidth]{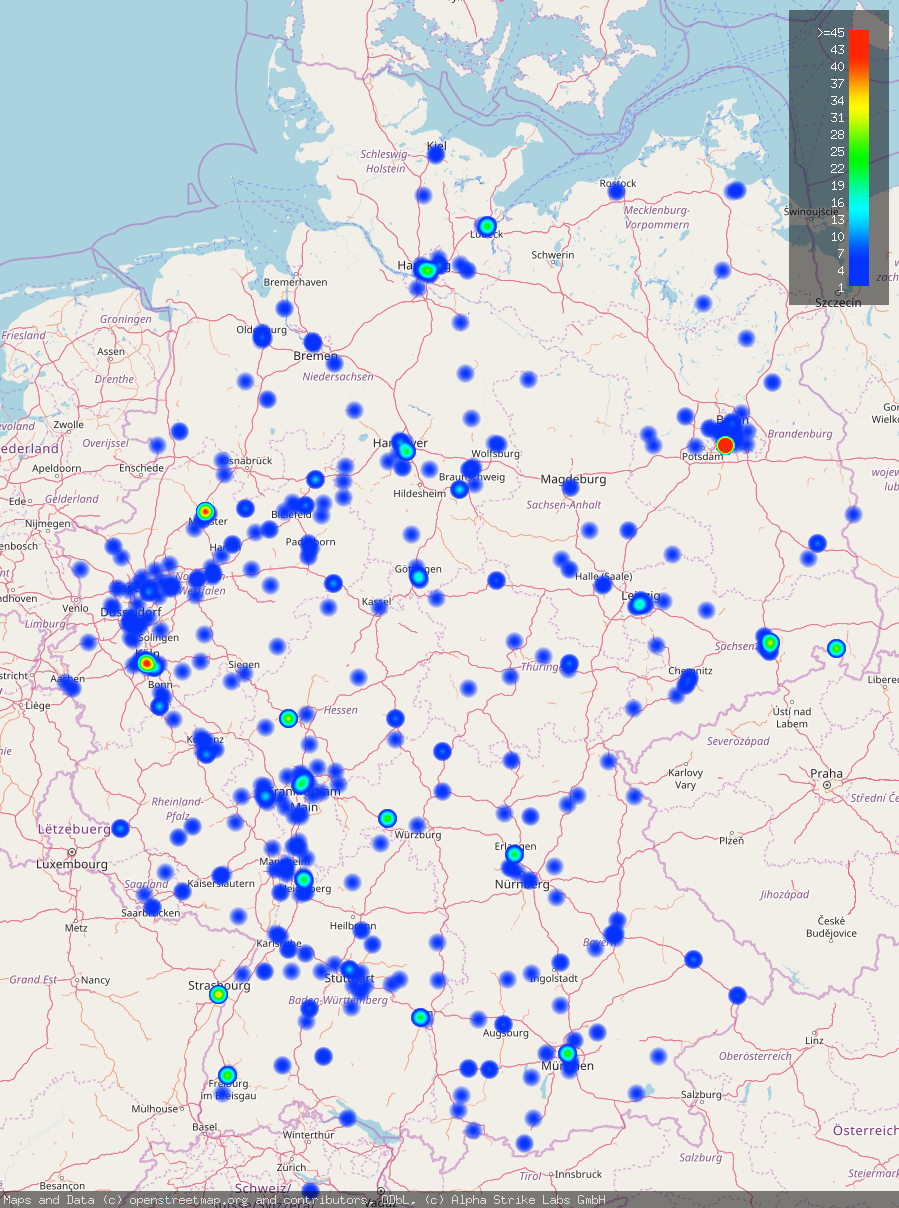}
    \caption{Geographical location of the network services with vulnerabilities as heatmap.} 
    \label{fig:cve}
  \end{subfigure} 
  \caption{Geolocation of hospitals, network services and vulnerabilities.}
  \label{fig:geoLoc}
\end{figure}

We created Figure \ref{fig:geoLoc} in order to show the geographic location of all 1,300 hospitals of the DIVI register. Here, it is easy to see that there is a high density of hospitals, particularly in the densely populated regions of western Germany and in the German metropolitan areas of Hamburg, Berlin and Munich (see Figure \ref{fig:divi}).  The image on the right (Figure \ref{fig:divi-vuln}) shows the DIVI registry hospitals with vulnerabilities on the map.  It is easy to recognize that both hospitals in metropolitan areas and in rural areas are affected.
\newpage

\newpage

In contrast to Figure \ref{fig:divi} and \ref{fig:divi-vuln}, Figure \ref{fig:geoIP} and \ref{fig:cve} represents an overview of all 1,555 identified hospitals and their 13,597 network services, which were assigned a geo-coordinate via a Geo-IP resolution. For the Geo-IP resolution, the commercial version of the Maxmind DB~\cite{maxmind} with increased resolution was used. 
Figure \ref{fig:geoIP} shows the network services of all hospitals analogous to Figure \ref{fig:divi}, whereas Figure \ref{fig:cve} only shows the network services with vulnerabilities.\\

The main difference between Figure \ref{fig:divi} and \ref{fig:cve} is that Figure \ref{fig:divi} only shows the hospitals of the DIVI registry and their geographical location. Figure \ref{fig:cve} however shows a heat map of all identified or vulnerable network services of German hospitals.
A comparison of the two graphs clearly reveals that the distribution in the heat map is somewhat smaller, but both graphs show that both rural regions and metropolitan areas have hospitals with vulnerabilities.\\



First of all the following overall CVSS vulnerability statistics should be noted:
\begin{table}[h!]
\begin{center}
\begin{tabular}{ c  c  }
  \hline
  CVSS-SCORE & Number of vulnerable services \\
  \hline
  9.0-10 (critical) & 931  \\
  7.0-8.9 (high) & 443  \\
  4.0-6.9 (medium) & 518  \\
  \hline
  Total vulnerable services: & 1,892 \\
   \hline
   \hline
\end{tabular}\\
\caption{CVSS distribution overview.}
\end{center}
\end{table}

Our analysis yielded overall 1,892 vulnerable services, with nearly half of the vulnerable services carrying a CVSS score of 9 or 10, thereby potentially containing critical vulnerabilities, according to version number.\\

\begin{figure}[htb!] 
  \begin{subfigure}[b]{0.98\linewidth}
    \centering
    \includegraphics[width=1\linewidth]{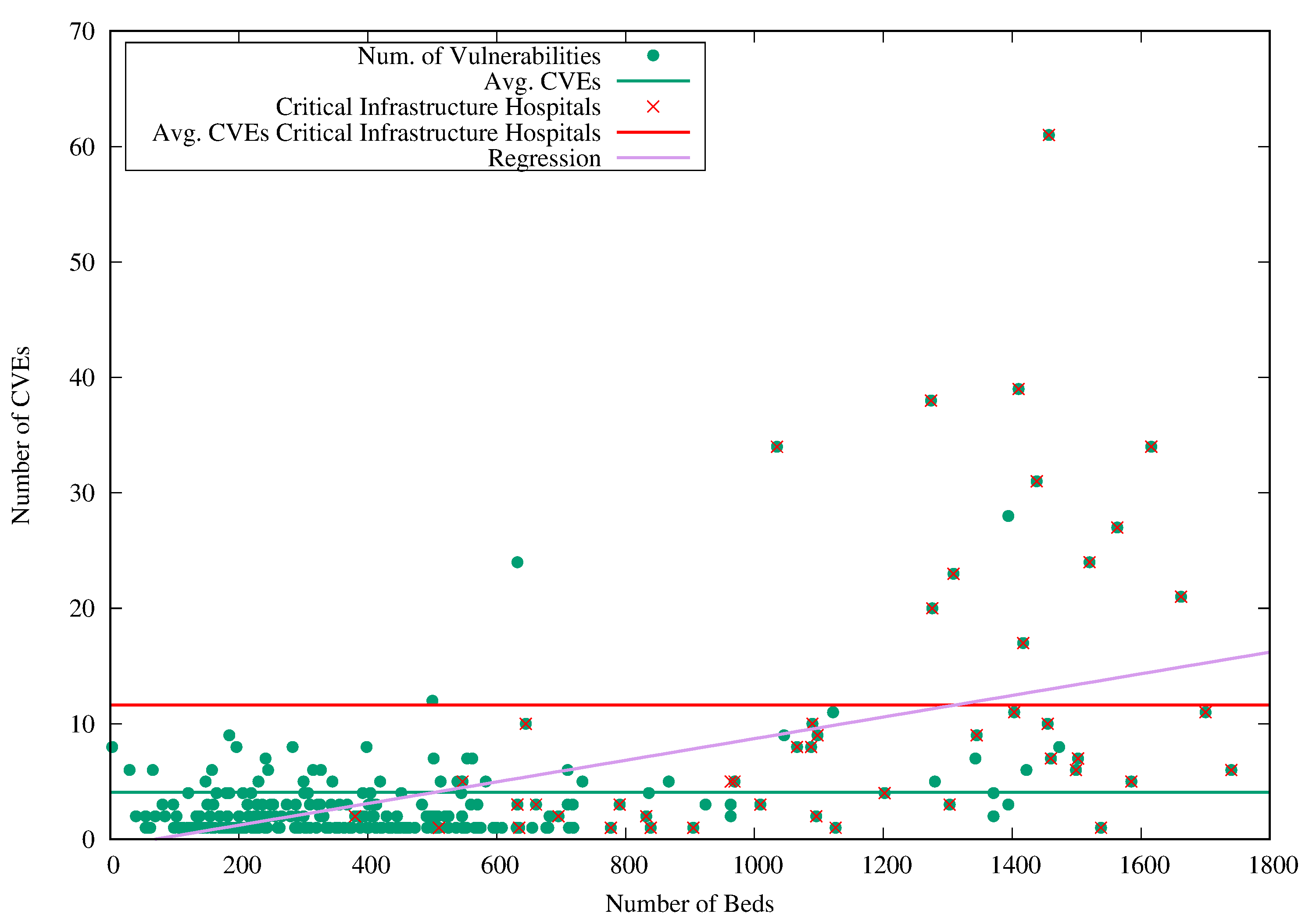}
    \caption{Number of vulnerabilities in hospitals with regard to the number of beds.} 
    \label{fig:BEDvsCVE}
  \end{subfigure} 
  \begin{subfigure}[b]{0.98\linewidth}
    \centering
    \includegraphics[width=1\textwidth]{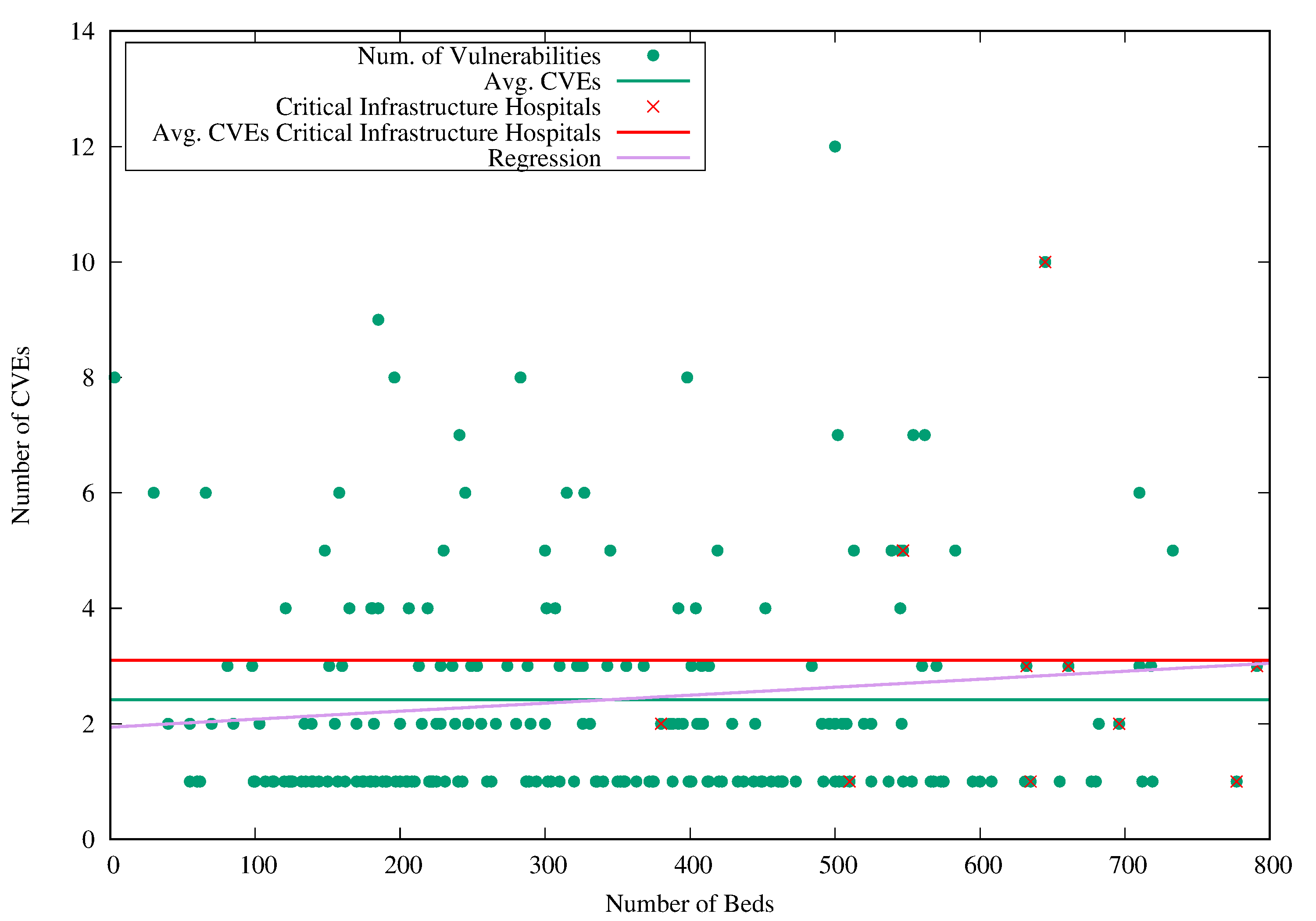}
    \caption{Detail view of the number of vulnerabilities in hospitals with up to 800 beds.}
    \label{fig:BEDvsCVEdetail} 
  \end{subfigure}
  \caption{Number of vulnerabilities in IT systems in hospitals.}
  \label{fig:NumVulnITHosp}
\end{figure}

Next we explore if there is any significance between the number of identified CVEs and the size of the clinical institution.
The allocation of the number of beds was taken from the German Hospital Register \cite{DKV21}.
An examination of the hospitals with vulnerabilities in relation to their bed capacity shows that about 167,000 
beds are hospitals with vulnerabilities. This represents 32 percent of the approx. 520,000 available hospital beds in Germany. 
Figure \ref{fig:NumVulnITHosp} shows the number of identified CVEs in relation to the size of the respective hospitals based on the number of beds.
For a better visualization, only hospitals with up to 1,800 beds are drawn; there are only a few facilities with a higher number of beds.\\

Since there is naturally a higher number of smaller hospitals, there are correspondingly more data points in the left-hand area of the figure. For better visibility, a detailed representation of this area is shown in Figure \ref{fig:BEDvsCVEdetail}.
\newpage



\newpage

A first look at the data initially reveals an unsurprising trend: As the number of beds increases, so does the number of vulnerabilities found in the IT environments of the respective hospitals.
This can probably be explained by the fact that larger hospitals with more beds also typically have more specialized medical departments and corresponding IT equipment, which thus increases not only the number of IT devices but also, in particular, the diversity of software and hardware used and thus the potential attack surface.
The regression line of this increase is drawn in the figures correspondingly.\\

But if we now look at the detailed view in Figure \ref{fig:BEDvsCVEdetail}, we notice that a corresponding increase in vulnerabilities in IT systems is much lower in the area of hospitals with up to 800 beds.\\

Hospitals with varying numbers of vulnerabilities occur here in all size ranges, without any particular characteristic being apparent.
This can be interpreted to imply that in the case of smaller hospitals, the number of existing vulnerabilities is more likely to depend on the quality of the respective IT service providers, or on specific software products.\\ 

With respect to the significantly increasing numbers of vulnerabilities at large hospitals however, especially those with bed numbers over 1,000, it is apparent that University hospitals in particular are to be found here more. This suggests that the higher CVE figures also reflect the need for more systems and, in the research sector, above all more diversified IT systems and scarcer or experimental software.\\

With respect to German legislation, the data in Figure \ref{fig:BEDvsCVE} offers yet another perspective for analysis:
Due to the special need for protection of the basic services necessary for modern society, such as electricity and water supply, telecommunications and healthcare, the BSI\footnote{Bundesamt für Sicherheit in der Informationstechnik (BSI), Federal Office for Information Security.} criticalness regulation (KRITIS Act~\cite{KritisV}) defines facilities in Germany that are obligated to implement minimum standards and measures in accordance with the BSI Act~\cite{BSIG} to ensure sufficient IT security.
In the area of hospitals, facilities with more than 30,000 in-patient cases per year are considered critical infrastructure.\\

Therefore, the interesting question arose whether the resulting liabilities are reflected in a lower visible CVE attack surface?\\

In order to evaluate this, the facilities that belong to  KRITIS based on the number of cases according to the German Hospital Register~\cite{DKV21} were marked accordingly in Figure \ref{fig:BEDvsCVE}.
Of course, large facilities such as University hospitals fall into this category, but so do some other, smaller facilities. Surprisingly, while the aforementioned accumulation of vulnerabilities can be seen at University hospitals, smaller institutions also feature systems with an above-average number of vulnerabilities.\\

To give an indication: For example, an average of 11.63 CVEs was identified for hospitals up to 1,800 beds belonging to KRITIS, while the average value for all of the hospitals up to 1,800 beds analyzed is 4.08.
A similar picture emerges when looking at the detail section of smaller hospitals in Figure \ref{fig:BEDvsCVEdetail}.
While the average number of CVEs present at the KRITIS hospitals is 3.1, all analyzed hospitals with up to 800 beds have an average of 2.42 CVEs.

\section{Discussion of results and conclusion}
\label{chap:discussion}

In our result section we firstly want to acknowledge the known limitations and constraints of our analysis, beginning with the number of 1,892 vulnerable services our DCS identified. Firstly, it must be noted that vulnerability identification is done fully automatically through service and banner mapping and CVE entry. In cases where patches have been backported or the administrator has changed banner information arbitrarily, the given CVE match indication naturally would not reflect the actual vulnerability state. Therefore, it may be assumed that the overall number might be a bit lower due to backports or banner changes. Secondly, although DCS uses a number of very well-proven port and service identification methods, in cases where fingerprinting fails this may create a situation where vulnerability identification is not always absolutely accurate.\\ 

Summarizing the results and findings: First of all it is quite noteworthy that looking at German clinical provider attack surface, analysis reveals many vulnerabilities also with quite high CVSS ratings. Looking at the most noteworthy system occurrences from a security point of view results are e.g. two Windows XP operating systems (CVSS 10.0 / End of Support since 2015!), open Jitsi VideoChat servers (CVSS 6.11), open unauthenticated squid proxies (CVSS 10.0) allowing proxy misuse, outdated Apache and PHP configurations (CVSS 9.8), direct accessible Intelligent Platform Management Interface (IPMI) Login Pages, Citrix XenAPP remote access (CVSS 10.0) or direct web links to RDP connections (CVSS 9.8) just to give a few concrete examples. The main required function of clinical institutions is healthcare and not IT security, however looking at the data, there still seems to be a need for better attack surface management, as approximately \textbf{32 percent of the analyzed services were determined as vulnerable to various degrees and 36 percent of all hospitals showed vulnerabilities.}\\

As mentioned, we can confirm that also healthcare institutions are affected to a certain extent by the issue of legacy services, for which end-of-support has been announced years ago and therefore security updates are not provided anymore.\\

Unsurprisingly, larger institutions have more IT systems, potentially leading to a larger attack surface, this was clearly visible in our analysis as well.\\

Finally, a rather interesting result of our analysis was the fact that hospitals belonging to German critical infrastructure, indicated through their assignment from the KRITIS Act, had \textbf{recognizable higher vulnerability number indicated by count of CVE numbers} in comparison to the entirety of analyzed hospitals. We found this striking, as our assumption was that KRITIS hospitals and clinics should have a better IT security posture, resulting in lower average CVE numbers in comparison with the rest of hospitals, as they are designated as being critical.\\

The findings on vulnerabilities at German hospitals underscore a key challenge in the KRITIS sector: a high number of outdated, sometimes proprietary systems that are difficult to patch, whether due to required re-certifications or end-of-support of software, combined with limited funding for IT security.\\
Our analysis concludes that even in 2020, despite its increased criticalness and increased regulation efforts, the German healthcare sector unfortunately presents and contains a certain visible amount of attack surface. This attack surface may translate into a national security risk, if abused systematically by an intelligent adversary. It is therefore advisable from a national risk management perspective, to regularly conduct reconnaissance in cyberspace on organizations that have been determined critical for a nation.



%
%

\printbibliography

\end{document}